\documentclass[fleqn,10pt]{wlscirep}
\usepackage[utf8]{inputenc}
\usepackage[T1]{fontenc}
\usepackage{multirow}
\title{Hybrid Attention Network for Accurate Breast Tumor Segmentation in Ultrasound Images}

\author[1,*]{Muhammad Azeem Aslam}
\author[2]{Asim Naveed}
\author[3]{Nisar Ahmed}
\affil[1]{School of Information Engineering, Xi'an Eurasia University, Xi'an, Shaanxi, 710071, China.}
\affil[2]{Department of Computer Science, University of Engineering and Technology Lahore, Faisalabad Campus, Faisalabad, 37630, Pakistan.}
\affil[3]{Department of Computer Science, University of Engineering and Technology Lahore, New Campus, Lahore, Punjab, 54890, Pakistan.}

\affil[*]{Corresponding author email: azeem@eurasia.edu}

\begin{abstract}
Breast ultrasound imaging is a valuable tool for early breast cancer detection, but automated tumor segmentation is challenging due to inherent noise, variations in scale of lesions, and fuzzy boundaries. To address these challenges, we propose a novel hybrid attention-based network for lesion segmentation. Our proposed architecture integrates a pre-trained DenseNet121 in the encoder part for robust feature extraction with a multi-branch attention-enhanced decoder tailored for breast ultrasound images. The bottleneck incorporates Global Spatial Attention (GSA), Position Encoding (PE), and Scaled Dot-Product Attention (SDPA) to learn global context, spatial relationships, and relative positional features. The Spatial Feature Enhancement Block (SFEB) is embedded at skip connections to refine and enhance spatial features, enabling the network to focus more effectively on tumor regions. A hybrid loss function combining Binary Cross-Entropy (BCE) and Jaccard Index loss optimizes both pixel-level accuracy and region-level overlap metrics, enhancing robustness to class imbalance and irregular tumor shapes. Experiments on public datasets demonstrate that our method outperforms existing approaches, highlighting its potential to assist radiologists in early and accurate breast cancer diagnosis.
\end{abstract}
\begin{document}

\flushbottom
\maketitle

\thispagestyle{empty}
\section{Introduction}
\label{sec:introduction}
Breast cancer remains one of the most prevalent and life-threatening malignancies among women globally \cite{zhang2024global}. Early and accurate detection is paramount to improving patient outcomes, where imaging-based modalities play a critical role in screening, diagnosis, and treatment planning \cite{zheng2023overview,karellas2008breast}. Among various imaging techniques, breast ultrasound imaging has emerged as a widely adopted complementary tool to mammography due to its non-ionizing nature, real-time imaging capabilities, cost-effectiveness, and high sensitivity in detecting tumors in dense breast tissues \cite{benson2004ultrasound,madjar2010role}. Unlike mammograms, which often struggle with overlapping tissue structures, ultrasound offers superior contrast for distinguishing solid masses from cystic structures and is especially beneficial for younger women with denser breast composition \cite{madjar2010role,gonzaga2010accurate}.\\
Despite its advantages, automated breast tumor segmentation in ultrasound images presents substantial challenges. Ultrasound images are inherently characterized by low signal-to-noise ratio, speckle noise, low contrast boundaries, and operator-dependent variability, which collectively hinder the reliable delineation of tumor margins \cite{liu2018automated}. Furthermore, intra-class variability and inter-class similarity between benign and malignant masses exacerbate the difficulty of precise segmentation, particularly in small datasets commonly encountered in medical imaging \cite{liu2018automated}. These challenges are less pronounced in mammographic images, where tissue structures are generally more consistent and the signal quality is higher \cite{[piqi],[GPR],ahmed2020perceptual}.\\
In recent years, deep learning-based methods \cite{ahmed2022deep,aslam2023vrl,aslam2024qualitynet,aslam2024tqp}, especially convolutional neural networks (CNNs), have demonstrated remarkable performance in medical image segmentation. Encoder-decoder architectures such as U-Net and its variants have become standard choices for biomedical applications due to their ability to capture both low-level textures and high-level semantics \cite{song2024survey}. However, traditional CNNs often fail to effectively capture long-range spatial dependencies and contextual information, which are critical for accurate segmentation in complex ultrasound data \cite{huang2024dra}. Moreover, simple skip connections used in many architectures may not be sufficient to propagate fine-grained details in the presence of severe noise and irregular tumor boundaries \cite{huang2024dra}.\\
To address these limitations, we propose a novel hybrid attention-based segmentation framework that integrates a pre-trained DenseNet121 encoder for robust feature extraction \cite{anari2025explainable} with a multi-branch attention-enhanced decoder tailored for breast ultrasound imagery. The architecture incorporates a combination of Global Spatial Attention (GSA) \cite{fang2020spatial}, Position Encoding (PE) \cite{murase2020can}, and Scaled Dot-Product Attention (SDPA) \cite{shen2022dilated} at the bottleneck, enabling the network to learn global context, spatial relationships, and relative positional features. In addition, Semantic Attention Blocks (SAB) \cite{fang2020spatial} are embedded at multiple decoding stages to refine and enhance spatial features, allowing the network to focus more effectively on tumor regions. This design facilitates superior localization and boundary preservation, crucial for clinical utility \cite{anari2025explainable,fang2020spatial}.\\
To optimize the learning process, we adopt a hybrid loss function combining Binary Cross-Entropy (BCE) and Jaccard Index Loss \cite{anari2025explainable}, balancing pixel-level accuracy with region-level overlap metrics, thereby making the model robust to class imbalance and irregular tumor shapes \cite{anari2025explainable,fang2020spatial,he2024sab}.\\
The key contributions of this work are summarized as follows:
\begin{itemize}
    \item We propose a DenseNet-based hybrid attention network tailored for robust tumor segmentation in ultrasound images.
    \item We introduce a contextual attention mechanism that combines spatial, positional, and semantic cues to improve segmentation accuracy.
    \item We incorporate a semantic attention block to reinforce decoder performance in reconstructing detailed and noise-tolerant segmentation maps.
    \item We employ a hybrid BCE and Jaccard loss to optimize both pixel-wise classification and mask-level similarity, addressing the challenges of speckle noise and structural ambiguity in ultrasound.
\end{itemize}
Through extensive experiments on publicly available breast ultrasound datasets, our proposed method demonstrates superior performance compared to existing state-of-the-art approaches, highlighting its potential for assisting radiologists in early and accurate breast cancer diagnosis. However, few models integrate spatial, semantic, and positional attention in a unified framework specifically tailored for the unique characteristics of ultrasound data. Our model effectively addresses this multi-dimensional attention gap by combining Global Spatial Attention (GSA), Scaled Dot-Product Attention (SDPA), and learnable Position Encoding (PE) within a hybrid architecture. Notably, this integration is achieved while maintaining low computational complexity, making the approach practical for real-time clinical deployment.
\section{Related Work}
The task of breast tumor segmentation in ultrasound (US) images has received substantial attention in recent years, primarily due to its clinical importance in facilitating early diagnosis and improving treatment planning \cite{zhang2024acl}. The inherent advantages of ultrasound over other imaging modalities—such as mammography—include its real-time imaging capability, lack of ionizing radiation, cost-effectiveness and better performance in dense breast tissues. However, these benefits come with unique challenges: ultrasound images often suffer from speckle noise, low contrast, operator dependency, and anatomical ambiguities, making automated segmentation significantly more challenging than in mammograms \cite{zhang2024acl,byra2020breast}. This has motivated the exploration of deep learning methods capable of extracting robust features and leveraging contextual information to overcome these limitations \cite{zhang2024acl,zhang2023fully}.
\subsection{Limitations of U-Net and CNNs}
Initial attempts at breast tumor segmentation relied on classical computer vision techniques such as filtering, active contours and clustering methods \cite{michael2021breast}. For example, threshold-based segmentation and graph-based approaches were used in early studies to delineate lesions in ultrasound images \cite{xu2024overview}. However, these methods required extensive domain knowledge and struggled with noise sensitivity and over-segmentation \cite{zhang2024acl,byra2020breast}.\\
The advent of convolutional neural networks (CNNs), particularly U-Net architectures, revolutionized the field by enabling end-to-end learning of hierarchical features \cite{zhang2024acl,zhang2023fully}. Recent studies demonstrate that densely connected U-Net variants with attention mechanisms achieve Dice scores exceeding 0.83, outperforming traditional methods \cite{zhang2024acl}. For instance, ACL-DUNet \cite{zhang2024acl} integrates spatial attention gates and channel attention modules to suppress irrelevant regions while enhancing tumor features. Similarly, SK-U-Net \cite{byra2020breast} employs selective kernels with dilated convolutions to adapt receptive fields, achieving a mean Dice score of 0.826 compared to 0.778 for standard U-Net.\\
To address limited contextual awareness, multi-branch architectures have emerged. One approach \cite{zhang2023fully} combines classification and segmentation branches, achieving a 0.991 AUC for normal/abnormal classification and a 0.898 Dice score for tumor segmentation. These models reduce false positives in normal images while maintaining sensitivity—a critical advancement for clinical screening \cite{zhang2023fully}. Hybrid designs like DeepCardinal-50 \cite{li2025efficient} further optimize computational efficiency, achieving 97\% accuracy in tumor detection with real-time processing capabilities.\\
Nevertheless, challenges persist in modeling long-range dependencies for lesions with irregular morphology. While attention mechanisms in ACL-DUNet improve spatial focus \cite{zhang2024acl}, and scale attention modules enhance multi-level feature integration \cite{zhang2024acl}, fuzzy boundaries in low-contrast ultrasound images remain problematic \cite{byra2020breast}. These constraints are being addressed by ongoing advancements in adaptive kernel selection and boundary-guided networks \cite{byra2020breast,zhang2024acl}.
\subsection{Rise of Attention Mechanisms}
Breast tumor segmentation in ultrasound imaging has seen significant advancements through the integration of attention mechanisms and use of hybrid architectures. Early approaches employed spatial-channel attention modules to address ambiguous boundaries and variable tumor sizes. For instance, SC-FCN-BLSTM \cite{pan2021tumor} combined bi-directional LSTM with spatial-channel attention to leverage inter-slice context in 3D automated breast ultrasound, achieving a Dice score of 0.8178. Abraham et al. \cite{abraham2019novel} explored hybrid attention mechanisms that effectively reweight feature maps based on contextual saliency, significantly improving segmentation quality in noisy ultrasound environments. Similarly, adaptive attention modules like HAAM \cite{chen2022aau} replaced traditional convolutions in U-Net variants, enabling dynamic receptive field selection across channel and spatial dimensions for improved segmentation robustness.\\
The CBAM-RIUnet \cite{benson2004ultrasound} further enhanced performance by integrating convolutional block attention modules with residual inception blocks, achieving Dice and IoU scores of 89.38\% and 88.71\%, respectively which is particularly effective in eliminating irrelevant features. Chen et al. \cite{chen2024esknet} proposed ESKNet by integrating selective kernel networks into U-Net to dynamically adjust the receptive field using attention mechanisms, enhancing performance across lesion types.\\
Although attention-based models have improved segmentation accuracy, many approaches still fail to adequately represent long-range spatial relationships, specifically when relying on a single attention strategy. This has led to the exploration of hybrid models that combine multiple attention mechanisms to provide a richer representation of both local and global features.
\subsection{Transformer-based Architectures in Medical Imaging}
Inspired by the success of transformers in natural language processing, vision transformers (ViTs) and their derivatives have begun to make significant inroads in medical image analysis \cite{xiao2023transformers}. Transformers enable global context modeling through self-attention mechanisms, overcoming the local limitations of CNNs. Several studies have successfully incorporated transformers into segmentation pipelines, either as standalone modules or in combination with CNN backbones \cite{liu2023optimizing,zhang2024automatic}.\\
In order to combine local convolutional features with long-range transformer-based context representations, He et al.\cite{he2023hctnet} and Ma et al.\cite{ma2023atfe} introduced hybrid CNN-transformer architectures.  Although these models work well, they frequently struggle to preserve fine-grained boundary information, which is essential for medical segmentation tasks. This is intended to be resolved by Swin Transformer-based networks, like DS-TransUNet \cite{lin2022ds}, which use shifted windows and hierarchical attention to capture both coarse and fine features at various scales. Similarly, Swin-Net \cite{zhu2024swin} use a swin-transformer encoder with feature refinement and enhancement module and hierarchical multi-scale feature
fusion module module to more precise segmentation. Whereas, SwinHR \cite{zhao2024swinhr} relies on hierarchical re-parameterization large kernel architecture to capture long-range dependencies while using shifted window-based self-attention mechanism for improved accuracy and computational efficiency.\\
Cao et al.\cite{cao2024neighbornet} took this further by developing a pixel-wise neighbor representation learning approach (NeighborNet), allowing each pixel to adaptively select its context based on local complexity. This approach is particularly suitable for ultrasound segmentation, where lesion boundaries may be fragmented or ambiguous.\\
In breast cancer segmentation, a critical research gap lies in the integration of transformer-based models with CNNs, where semantic mismatches between locally extracted CNN features and globally contextualized transformer representations often lead to suboptimal fusion \cite{zhang2024hau,wu2024mfmsnet}. Inflexible or disjointed fusion strategies, such as rigidly inserting transformer blocks into CNN architectures without addressing feature consistency, result in redundant or insufficient hierarchical representations \cite{wu2024mfmsnet}. This challenge is exacerbated in noisy or irregular data, such as breast ultrasound images, where speckle artifacts, shadowing, and blurred lesion boundaries create discordance between local texture details and global anatomical structures \cite{wu2024mfmsnet,tagnamas2024multi}. Current approaches frequently fail to bridge the semantic gap between CNN's localized feature extraction and transformers' long-range dependency modeling, particularly in decoder stages where misaligned feature maps reduce segmentation precision for small lesions and complex margins \cite{he2023hctnet}. Furthermore, the lack of adaptive cross-attention mechanisms to harmonize multi-scale features often diminishes model robustness against ultrasound-specific noise patterns \cite{wu2024mfmsnet}, highlighting the need for more sophisticated hybrid architectures that enable synergistic local-global feature interaction while maintaining computational efficiency \cite{he2023hctnet}.
\subsection{Gap in Hybrid Attention Design}
Our framework advances breast ultrasound segmentation by introducing a novel hybrid attention-based network. This network strategically integrates a DenseNet121 encoder (pre-trained on ImageNet) with transformer-inspired attention mechanisms (GSA, SDPA), and learnable position encoding (PE) to address semantic mismatches between local CNN features and global context representations. Addressing insufficient boundary preservation, Semantic Attention Blocks (SAB) are incorporated into the decoder pathway, dynamically reconciling multi-scale features through context-guided channel reweighting; bilinear upsampling is employed to reduce checkerboard artifacts compared to transposed convolutions. To ensure robustness, the network is trained using a hybrid BCE + Jaccard loss function explicitly designed to handle ultrasound-specific noise and morphological irregularities, features often absent in current models.
\section{Methodology}
\begin{figure}[hbtp]
    \centering
    \includegraphics[width=0.8\textwidth]{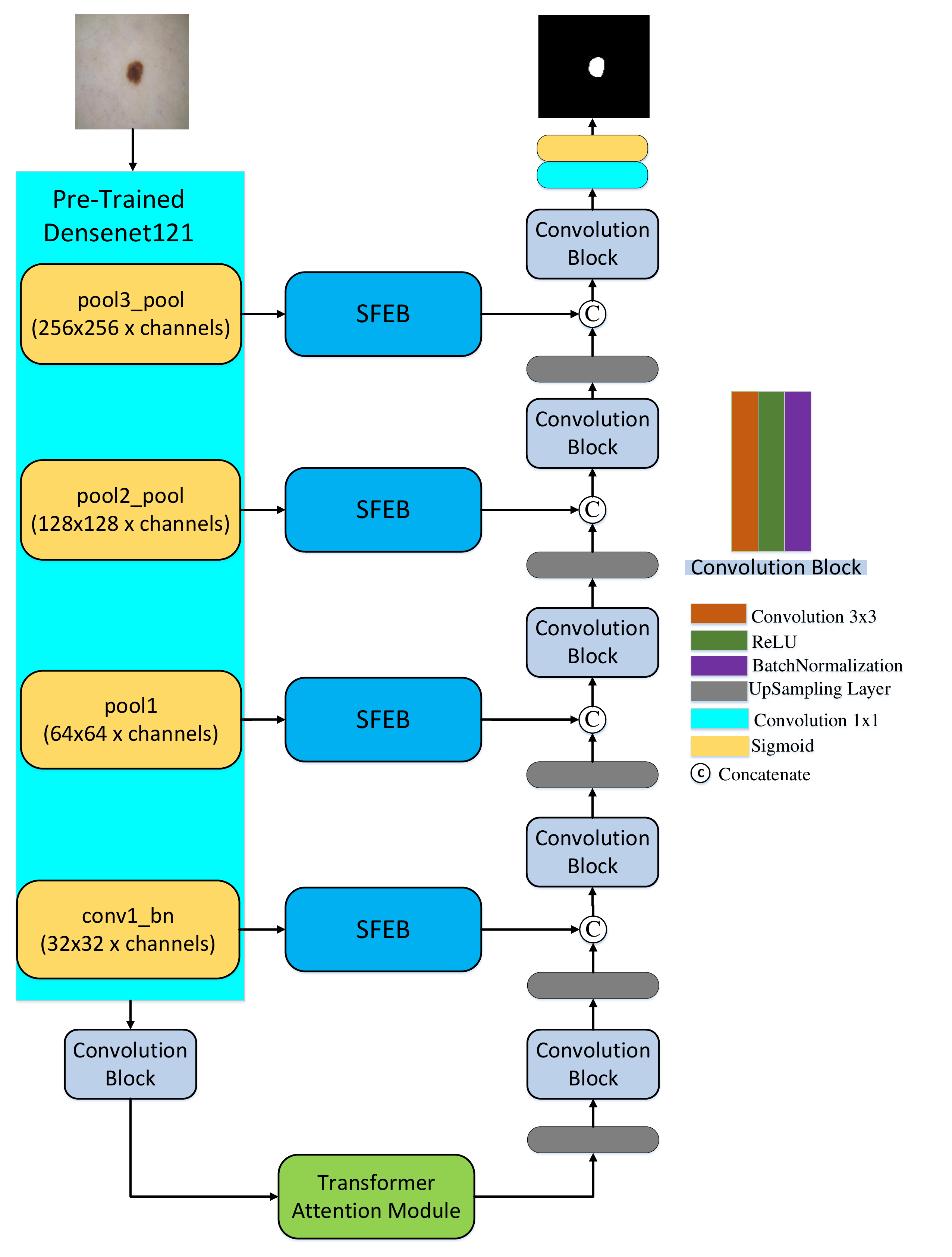}
    \caption{The details of the proposed method. The proposed method consists of a pre-trained encoder and a specific decoder, a spatial features enhancement block (SFEB), and a Transformer attention module.}
    \label{model}
    % \vspace{0.5cm}
\end{figure}

The proposed method is structured into four key stages: encoding, decoding, Transformer-based attention, and a spatial feature enhancement block. To efficiently extract complex image features, the model employs DenseNet121 \cite{huang2017densely} as its encoder backbone. DenseNet’s architecture enhances feature representation by establishing direct connections between layers within dense blocks (DBs), promoting feature reuse, and ensuring seamless information flow. These characteristics are particularly beneficial in medical image analysis, where capturing intricate details is crucial for precise segmentation.\\
The encoding stage consists of four sequential encoding blocks, each comprising four dense blocks (DBs) followed by three transition layers (TLs), as illustrated in the Figure~\ref{model}. This hierarchical design allows the model to progressively extract diverse and meaningful features from medical images. The DB and TL configurations further enhance the model’s ability to capture critical image details throughout the encoding process, ensuring a rich and robust feature representation.
The decoding stage follows a lightweight approach inspired by the U-Net \cite{ronneberger2015u}, but with a reduced number of parameters to enhance efficiency. Instead of standard transposed convolutions, which can introduce checkerboard artifacts and increase computational overhead, our method employs bilinear upsampling followed by lightweight convolutional layers. This strategy maintains segmentation accuracy while significantly reducing the model's computational cost.\\
Additionally, skip connections from the encoder are incorporated to retain spatial details and enhance feature fusion at different resolution levels. By integrating the spatial feature enhancement block (SFEB) within the decoding pathway, the model further refines feature maps, improving segmentation accuracy with fewer parameters compared to a standard U-Net. Overall, this optimized design ensures efficient processing of medical images while preserving fine structural details, making the model well-suited for real-world clinical applications.

\begin{figure*}[ht]
\centering
     \includegraphics[scale=0.65]{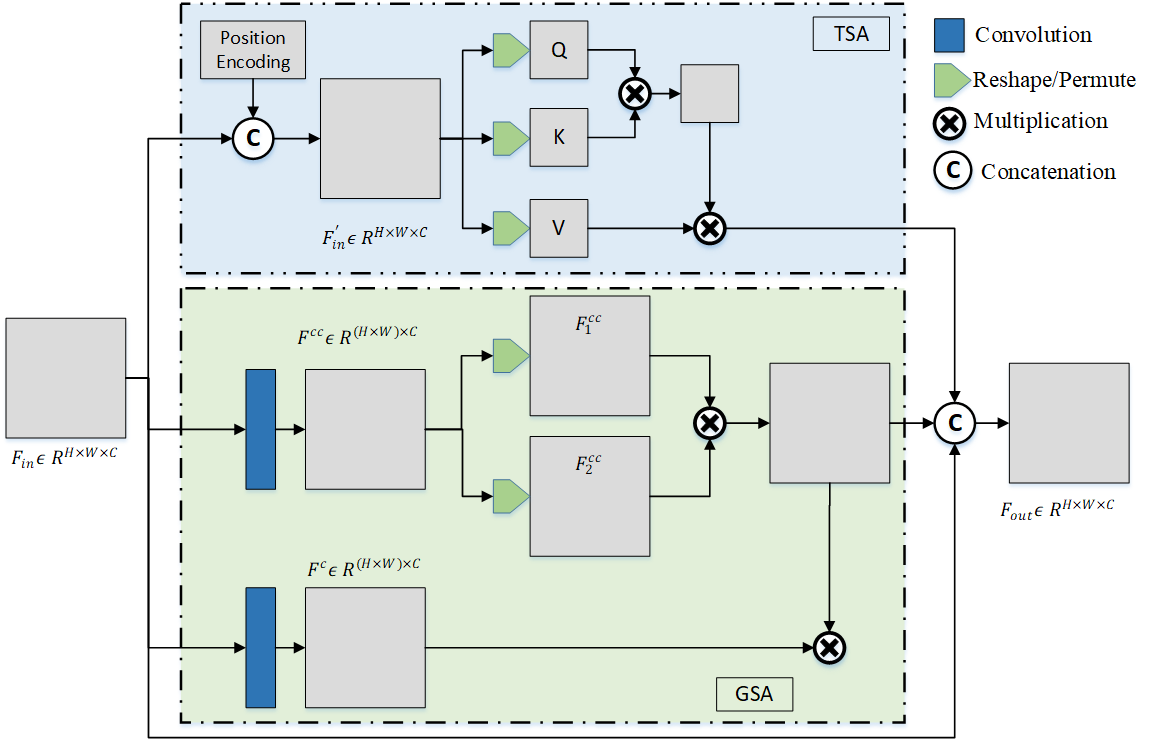}
    \caption{The details of the self-aware attention module. The top block shows the transformer self-attention, and the bottom shows the global self-attention block.}
    \label{transformer}
    % \vspace{0.5cm}
\end{figure*}

\subsection{Transformer Attention Module (TAM)} 
To enhance the model’s ability to capture and integrate contextual information, we incorporate a self-aware attention module \cite{chen2023transattunet}. This module consists of two key components. Firstly, the Transformer Self-Attention (TSA) block captures contextual information by considering relative positions within the input data. It integrates positional information by concatenating input features with positional embeddings, allowing the model to better understand spatial relationships within the input data. Secondly, the Global Spatial Attention (GSA) block refines local contextual information by aggregating it with global features. By incorporating a broader perspective, it enhances the model’s ability to retain fine details while maintaining a comprehensive understanding of the overall structure.
Collectively, these attention mechanisms improve feature representation, helping the model effectively balance local and global information for more precise segmentation.
\subsubsection{Transformer Self-Attention (TSA)}
Since multi-head attention effectively captures self-correlation but cannot learn spatial relationships, a common strategy is to introduce positional encoding before applying attention. Specifically, the input feature map $F\in \mathbb{R}^{h\times w\times c}$ is first enriched with positional information, producing a representation, which is then fed into the multi-head attention block (Fig.~\ref{transformer}). $F$ is projected into three distinct matrices using learnable weights, including the matrix of queries $Q\in \mathbb{R}^{ c \times (h\times w)}$, the matrix of keys $K\in \mathbb{R}^{ c \times (h\times w)}$, and $ V \in \mathbb{R}^{h\times w\times c}$. 

\begin{equation}\label{}
Q = F\cdot W_{q}
\end{equation}
\begin{equation}\label{}
K = F\cdot W_{k}
\end{equation}
\begin{equation}\label{}
V = F\cdot W_{v}
\end{equation}
where the $W_{q}, W_{v}, W_{k}$ are the embedding matrices for different linear projections.
The scaled dot-product attention mechanism computes the similarity between different channels by applying the Softmax-normalized dot product of $Q$ and the transposed version of $K$. This matrix gives the contextual attention map $ A \in \mathbb{R}^{c\times c}$. \\
Finally, by multiplying the contextual attention map $A$ with $V$, attention-weighted feature representations are obtained. This process enables the multi-head attention mechanism to refine feature aggregation while preserving crucial contextual dependencies. The overall formulation can be expressed as:
\begin{equation}\label{}
A_{TSA}{(Q,K,V)} = softmax(\frac{QK^T}{\sqrt d_k})V,
\end{equation}
In this way, we optimize the shape of the output feature maps. 
\subsubsection{Global Spatial Attention (GSA)}
The TAM uses the GSA component to selectively aggregate global context with learned attributes and encode larger information.
Incorporating contextual positioning information into local features enhances intra-class compactness and optimizes feature representation. Figure 2 depicts the architecture of global spatial attention. The input feature map $F\in \mathbb{R}^{h\times w\times c}$
is embedded in $F^c\in \mathbb{R}^{h\times w\times c}$ and $F^{cc}\in \mathbb{R}^{h\times w\times c'}$  where c' = c/2. The latter is reshaped to $F1^{cc}\in \mathbb{R}^{h\times w\times c'}$ and $F2^{cc}\in \mathbb{R}^{h\times w\times c'}$, their scaled dot product is computed and passed through a Softmax normalization layer, producing an attention map $GSA\in \mathbb{R}^{(h\times w) \times (h\times w)}$, which represents the correlation between different spatial positions.
The multi-head attention mechanism is then formulated as:
\begin{equation}\label{}
A_{GSA} = softmax(F1^{cc} \cdot F2^{cc})
\end{equation}
Finally, the output feature map ($F_{out}\in \mathbb{R}^{h\times w\times c}$) of the self-aware attention module is obtained by concatenating the outputs from Transformer Self-Attention (TSA), Global Spatial Attention (GSA), and the original input. This approach ensures that both global context and local spatial dependencies are effectively captured, enhancing the model’s ability to extract meaningful features for accurate segmentation.
\subsection{Spatial Features Enhancement Block (SFEB)}
\begin{figure*}[ht]
\centering
     \includegraphics[scale=0.68]{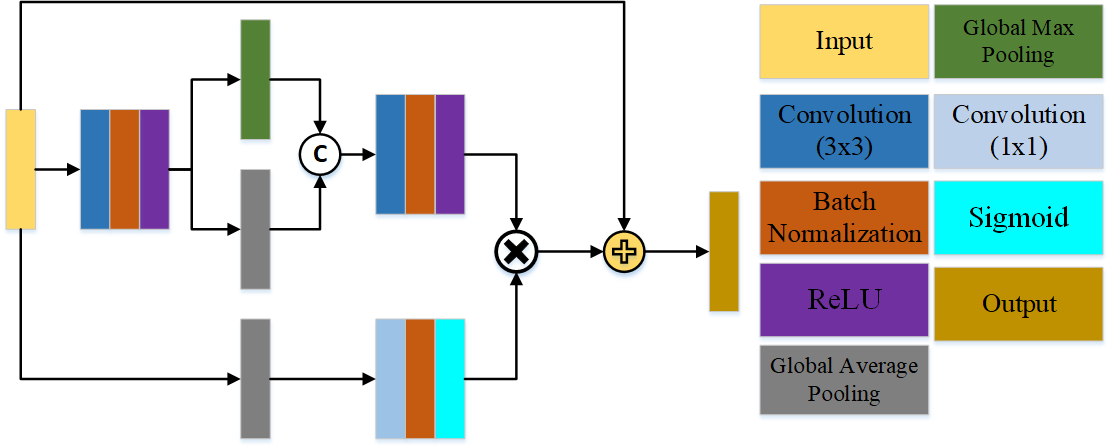}
    \caption{Spatial Features Enhancement Block}
    \label{SFEB}
    % \vspace{0.5cm}
\end{figure*}

Pooling operations play a crucial role in deep learning models by reducing the size of feature maps, speeding up computations, and improving feature robustness. When it comes to lesion segmentation, it's essential to capture both fine details and broader contextual information. This is because lesions often appear in varying colors, have low contrast, and are generally small in size. To address these challenges, our approach integrates a Spatial Feature Enhancement Block (SFEB) within skip connections to improve feature fusion, attention mechanisms, spatial information processing, and residual learning. These enhancements contribute to more precise and resilient segmentation, particularly when maintaining fine details and preserving object boundaries is critical.\\
The input tensor undergoes several processing steps, starting with a $3 \times 3$ convolutional layer, followed by batch normalization (BN) and a Rectified Linear Unit (ReLU) activation. To capture global feature information, both global max-pooling and global average-pooling are applied, and their outputs are combined. This fusion strengthens the model’s ability to extract meaningful global features, leading to more accurate predictions. The concatenated feature maps are then passed through another $3 \times 3$ convolutional layer, BN, and ReLU, refining the extracted features for improved segmentation accuracy.\\
Moreover, a parallel path is introduced, incorporating global average pooling, a $1 \times 1$ convolutional layer, BN, and a sigmoid activation function. This path generates attention coefficients that adjust the importance of the pooled features before integrating them with the original input. By leveraging both max-pooling and average-pooling in parallel, along with attention-based weighting, the SFEB architecture effectively balances local and global feature information. This strategy enhances the model’s capability to perform high-accuracy lesion segmentation. The details of the SFEB are shown in Figure~\ref{SFEB}.\\
The SFEB technique, as demonstrated in the following equations, strengthens the representation of features in the given input tensor, ultimately improving the model's segmentation performance.
\begin{equation}\label{Eq1}
{I} =  \mathbb{R}^{H\times W\times C}
\end{equation}
In Equation \ref{Eq1}, the symbol $I$ represents input, where $H$ denotes height, $W$ represents width, and $C$ signifies its depth, resulting in dimensions $H \times W \times C$. These dimensions define the size and depth of the input tensor.
\begin{equation}\label{Eq2}
I_{1} = \text{ReLU}(\mu(f^{3\times3}(I)),
\end{equation}
Equation \ref{Eq2} represents the output $I_1$, which is acquired by convolving the input tensor $I$ with a $3 \times 3$ filter ($f^{3\times3}$), afterward batch normalization ($\mu$) and $ReLU$.

\begin{equation}\label{Eq3}
GP_{m} = (G_{m}(I_1)),
\end{equation}
\begin{equation}\label{Eq4}
GP_{a} = (G_{a}(I_1)),
\end{equation}

Equations \ref{Eq3} and \ref{Eq4} represent $ GP_m$ and $GP_a$, respectively, which are obtained by applying the global maximum-pooling operation ($G_m$) and the global average-pooling ($G_a$) operation on $I_1$.

\begin{equation}\label{Eq5}
P_{o} = GP_{m} \copyright GP_{a},
\end{equation}
In Equation \ref{Eq5}, $P_{o}$ is obtained by concatenating the global maximum-pooling ($G_m$) and global average-pooling ($G_a$) features.

\begin{equation}\label{Eq6}
F_{c} = \text{ReLU}(\mu(f^{3\times3}(P_{o})),
\end{equation}

Equation \ref{Eq6} shows $F_c$, obtained by applying a convolution operation ($f^{3\times3}$), subsequently batch normalization ($\mu$) and ReLU activation ($ReLU$) on complementary characteristics $P_o$.

\begin{equation}\label{Eq7}
F_{G} = GAP(I),
\end{equation}

In Equation \ref{Eq7}, $F_G$ is derived by performing a global average pooling operation ($GAP$) on the input tensor $I$.

\begin{equation}\label{Eq8}
F_{cc} = \sigma(\mu(f^{1\times1} (F_{G}))),
\end{equation}

Equation \ref{Eq8} shows $F_{cc}$, obtained by applying a convolution operation ($f^{1\times1}$), subsequently ($\mu$) and sigmoid activation ($\sigma$) in $F_G$.

\begin{equation}\label{Eq9}
F_{em} = F_{c}\otimes F_{cc},
\end{equation}
$F_{em}$ is computed through element-wise multiplication ($\otimes$) of $F_c$ and $F_{cc}$, as described in Equation \ref{Eq9}.

\begin{equation}\label{Eq10}
{F} = F_{em}\oplus I,
\end{equation}
Ultimately, the attention features $F$ are acquired by element-wise summation ($\oplus$) of $F_{em}$ and the input tensor $I$, as delineated in Equation \ref{Eq10}.

\subsection{Loss Functions}
For the optimized training of deep learning models, understanding the loss functions is extremely important \cite{jadon2020survey}.
% \textbf{Binary cross entropy loss:}
Binary cross-entropy loss is used to evaluate each projected probability against the class output.
Eq. \ref{bce} explains it mathematically \cite{jadon2020survey}
\begin{equation}\label{bce}
\mathrm{Loss}_\mathrm{bce} = -\sum_{i=1}^{N} ({y}_{i}\mathrm{log}{\hat{y}}_{i}  + (1-{y}_{i})\mathrm{log}(1-{\hat{y}}_{i})).
\end{equation}
where the model output value is computed by ${\hat{y}}_{i}$ and the corresponding value is ${y}_{i}$,
% \textbf{Dice loss:}

% \textbf{Jaccard loss:}
Jaccard (IOU) loss is a region-based loss and calculates the intersection over the union between two images \cite{van2019deep}
\begin{equation}\label{jaccard}
\mathrm{Loss}_\mathrm{jaccard} = 1-\frac{\sum_{i}^{N}(P_{i}G_{i})}{\sum_{i}^{N}(P_{i} + G_{i} - P_{i}G_{i}) }.
\end{equation}
The combined loss function is used for the evaluation of the proposed method as mentioned in Eq.~\ref{combined}. 
\begin{equation}\label{combined}
\mathrm{Loss}_\mathrm{combined} = \mathrm{Loss}_\mathrm{bce} + \mathrm{Loss}_\mathrm{jaccard} .
\end{equation}

\section{Experiments}

\subsection{Datasets for Breast Ultrasound Image Segmentation}
\label{subsec:datasets}
To evaluate the effectiveness of the proposed breast cancer segmentation approach, we utilized two publicly available breast ultrasound datasets: BUSI and UDIAT. Both datasets provide B-mode grayscale ultrasound images accompanied by expert annotations, offering a suitable benchmark for tumor segmentation tasks.\\
\textbf{BUSI Dataset:} The Breast Ultrasound Image (BUSI) dataset \cite{al2020dataset} contains 780 grayscale ultrasound images acquired from 600 female patients aged between 25 and 75 years. Each image has an approximate resolution of $500 \times 500$ pixels and is categorized into one of three classes: normal, benign, or malignant. For our segmentation experiments, only the benign and malignant images were retained, as these include corresponding binary segmentation masks delineating tumor regions annotated by clinical experts. The normal class images were excluded due to the absence of tumors. The images were resized to $256 \times 256$ for uniform processing during model training and evaluation.\\
\textbf{UDIAT Dataset:} The UDIAT dataset, compiled by Yap et al. \cite{yap2020breast}, comprises 163 B-mode breast ultrasound images acquired using a Siemens ACUSON Sequoia C512 diagnostic ultrasound system. The average resolution of the images is $760 \times 570$ pixels. Similar to BUSI, each image in the UDIAT dataset is labeled as either benign or malignant and is provided with corresponding pixel-wise tumor segmentation masks. The images were similarly resized to $256 \times 256$ pixels for consistency with the BUSI dataset.\\
The division of each dataset into training and test sets is summarized in Table~\ref{datasets}. No separate validation set was used; cross-validation or hold-out test evaluation strategies were employed as described in subsequent sections.
\begin{table*}[htbp]
  \centering
  \caption{Details of the datasets used in this study.}
    \begin{tabular}{ccccc}
    \hline
    \multirow{2}[4]{*}{\textbf{Dataset}} & \multicolumn{3}{c}{\textbf{Number of Images}} & \multirow{2}[4]{*}{\textbf{Image Resolution}} \\
\cmidrule{2-4}          & \textbf{Training} & \textbf{Test} & \textbf{Validation} &  \\
    \hline
    BUSI  & 700   & 80    & -     & $256 \times 256$ \\
    UDIAT & 133   & 33    & -     & $256 \times 256$ \\
    \hline
    \end{tabular}%
  \label{datasets}%
\end{table*}%
\subsection{Implementation Details}
\label{subsec:implementation}

All input images from both datasets were resized to a uniform resolution of $256 \times 256$ pixels to ensure compatibility with the model architecture. During training, 20\% of the training data was reserved as a validation set to facilitate performance monitoring and hyperparameter tuning. Model optimization was carried out using the Adam optimizer \cite{kingma2014adam}, initialized with a learning rate of 0.001. The learning rate was adaptively reduced by a factor of 0.25 if no improvement was observed on the validation set for four consecutive epochs. Early stopping was employed to mitigate overfitting and automatically terminate training once convergence was achieved.\\
A hybrid loss function was used to guide the optimization process, and a batch size of 10 was selected for training. Notably, the model achieved strong performance without the application of any data augmentation techniques.\\
All experiments were implemented using the Keras framework with TensorFlow as the backend. The training and evaluation procedures were executed on an NVIDIA Tesla K80 GPU, supported by an Intel Xeon 2.20 GHz CPU, 13 GB of system RAM, and 12 GB of GPU memory.
\subsection{Evaluation Metrics}
\label{subsec:evaluation}
To quantitatively evaluate the segmentation performance of the proposed method, a set of widely accepted metrics in medical image analysis was employed. These metrics provide a comprehensive assessment of both pixel-wise classification accuracy and region-level overlap between the predicted segmentation masks and the ground truth annotations.\\
\textbf{Jaccard Index (Intersection over Union, IoU)}
The Jaccard Index measures the degree of overlap between the predicted and ground truth segmentation masks. It is defined as:
\begin{equation}
\mathrm{IoU} = \frac{TP}{TP + FP + FN}
\end{equation}
where $TP$, $FP$, and $FN$ denote the number of true positive, false positive, and false negative pixels, respectively. A higher IoU indicates a greater alignment between the predicted and actual tumor regions.
\textbf{Accuracy}
This metric represents the overall proportion of correctly classified pixels, including both positive (tumor) and negative (non-tumor) classes:
\begin{equation}
\mathrm{Accuracy} = \frac{TP + TN}{TP + TN + FP + FN}
\end{equation}
\textbf{Recall (Sensitivity)}
Recall quantifies the model’s ability to correctly identify tumor pixels:
\begin{equation}
\mathrm{Recall} = \frac{TP}{TP + FN}
\end{equation}
It is particularly critical in medical diagnostics, where minimizing false negatives is essential.
\textbf{Precision}
Precision reflects the proportion of correctly predicted tumor pixels among all pixels predicted as tumor:
\begin{equation}
\mathrm{Precision} = \frac{TP}{TP + FP}
\end{equation}
\textbf{Dice Similarity Coefficient (DSC)}
The Dice score is a widely used metric in segmentation tasks, especially for imbalanced datasets. It is the harmonic mean of precision and recall:
\begin{equation}
\mathrm{Dice} = \frac{2 \times TP}{2 \times TP + FP + FN}
\end{equation}
The Dice coefficient is particularly informative when the region of interest occupies a small fraction of the image.
\textbf{Specificity}
Specificity measures the ability of the model to correctly classify background (non-tumor) pixels:
\begin{equation}
\mathrm{Specificity} = \frac{TN}{TN + FP}
\end{equation}
High specificity ensures that normal regions are not mistakenly identified as lesions, which is crucial for reducing false alarms in clinical practice.
\subsection{Ablation Studies}
\label{subsec:ablation}
We conducted a series of ablation experiments on the BUSI dataset to evaluate the contribution of each key component within the proposed segmentation framework. The encoder backbone of the model is based on a pre-trained DenseNet-121, chosen for its strong feature extraction capabilities. To enhance spatial representation and contextual understanding, we progressively integrated a Convolutional Block, a Spatial Enhancement Feature Block (SEFB), and a Transformer-based Attention Module (TAM) into the baseline architecture.\\
The quantitative results presented in Table~\ref{tab:ablationresults} demonstrate the effectiveness of these components. Starting from the baseline configuration, each successive addition yields consistent improvements across all evaluation metrics. In particular, the inclusion of SEFB and TAM significantly boosts segmentation performance, as reflected in the Jaccard index, Dice coefficient, sensitivity, precision, and overall accuracy. These findings confirm the critical role of each module in enhancing the model’s capability to accurately delineate tumor regions.
\begin{table*}[htbp]
\centering
\caption{Performance of ablation experiments on the BUSI dataset.}
\begin{tabular}{lccccccc}
\hline
\textbf{Method} & \textbf{Parameters} & \textbf{Jaccard} & \textbf{Dice} & \textbf{Sensitivity} & \textbf{Accuracy} & \textbf{Precision} & \textbf{Specificity} \\
\hline
U-Net \cite{ronneberger2015u} & 34,513,345 & 76.54 & 83.13 & 82.83 & 97.91 & 83.94 & 98.81 \\
Baseline (BL) & 10,498,945 & 90.44 & 94.37 & 94.33 & 99.46 & 94.43 & 99.69 \\
BL + ConvBlock & 11,089,537 & 92.89 & 96.27 & 96.33 & 99.60 & 96.44 & 99.77 \\
BL + ConvBlock + SEFB & 15,339,329 & 93.51 & 96.80 & 96.58 & 99.69 & 96.66 & 99.82 \\
BL + ConvBlock + SEFB + TAM & 15,427,713 & \textbf{94.75} & \textbf{97.28} & \textbf{97.15} & \textbf{99.74} & \textbf{97.42} & \textbf{99.84} \\
\hline
\end{tabular}
\label{tab:ablationresults}
\end{table*}
% 4. Experiments and Results
% 4.1 Implementation Details
% •	Hardware and software specifications
% •	Hyperparameter tuning (learning rate, batch size, epochs)
% 4.2 Evaluation Metrics
% •	Explanation and formulas for metrics used (Dice, Jaccard, Sensitivity, Specificity, Accuracy, Precision)
% 4.3 Experimental Results
% 4.3.1 Ablation Studies
% •	Systematic analysis of each proposed component (tables and figures)
% 4.3.2 Comparative Analysis
% •	Performance comparison with state-of-the-art methods
% •	Qualitative visualization results (segmentation output examples)
% 5. Discussion
% 5.1 Model Performance Analysis
% •	Detailed interpretation of quantitative results
% •	Strengths and improvements compared to existing methods
% 5.2 Limitations of the Proposed Approach
% •	Clear acknowledgment and discussion of limitations
% •	Potential scenarios where performance may vary
% 5.3 Clinical Implications
% •	Practical impact of your proposed method on clinical practices
% •	Potential for real-world deployment and integration
% 6. Conclusion
% •	Concise restatement of research objectives and findings
% •	Key strengths of the proposed approach
% •	Recommendations for future research directions
\section{Results and Discussion}
\subsection{Comparison with State-of-the-Art Methods on the BUSI Dataset}
\label{subsec:results_busi}
To rigorously evaluate the effectiveness of the proposed segmentation framework, we conducted comprehensive comparisons with a diverse set of state-of-the-art (SOTA) deep learning models on the BUSI breast ultrasound dataset. The selected models represent recent advances in medical image segmentation and include BGRA-GSA, AAU-Net, MLFEU-Net, Swin-UNet, Eh-Former, U-Net, BGRD-TransUNet, Attention U-Net, UNet++, and DDRA-Net.\\
As summarized in Table~\ref{tab:busiresults}, the proposed method consistently outperforms all competing approaches across multiple evaluation metrics, including Dice Similarity Coefficient (DSC), Jaccard Index (IoU), sensitivity, precision, specificity, and accuracy. These results highlight the robustness of our approach in segmenting complex breast ultrasound images, which often suffer from challenges such as low contrast, speckle noise, and ambiguous lesion boundaries.
\begin{table*}[htbp] \centering
  \caption{Performance comparison with state-of-the-art methods on the BUSI dataset.}
    % \adjustbox{max width=\textwidth}{
    \begin{tabular}{lccccccc}
    \hline
     \textbf{Method}  & \textbf{$J_{i}$} & \textbf{$D_{c}$}  & \textbf{$S_{n}$} & \textbf{$A_{cc}$} & \textbf{$P_{r}$} & \textbf{$S_{p}$} \\
    \hline 
    BGRA-GSA \cite{hu2023boundary} & 68.75 & 81.43&84.14 & 96.34 & 79.01 & 97.63\\
    AAU-Net \cite{chen2022aau} & 69.26 & 78.18 & 86.06 & - & 81.17 &99.17\\
    MLFEU-NET \cite{tang2025mlfeu} &  71.86 & 80.82 & 82.80 & - & 85.22 & 99.46 & \\
    Swin-unet \cite{cao2023swin} &74.16& 79.45& 83.16 & 96.55& -&97.34\\
    Eh-former \cite{qu2024eh} & 76.37 & 84.6 & 87.74 &- &- & 98.17\\
    U-Net \cite{ronneberger2015u} &76.54 & 83.13 & 82.83 & 97.91 & 83.94 & 98.81\\
    BGRD-TransUNet \cite{ji2024bgrd} & 76.77 & 85.08 & 87.62 & 97.14 & 85.89 & -\\
    Attention U-Net \cite{oktay2018attention} & 77.89 & 85.96 & 85.80 & 97.85& 86.65& 98.54\\
    Unet++ \cite{zhou2018unet++}  & 81.09 & 88.11 & 87.29 & 98.57 & 89.53 & 99.18\\
    DDRA-Net \cite{sun2024ddra} &89.23&75.32&92.32&-&95.02&-\\
    
    \hline
    % \hdashline
    Proposed method  & \textbf{94.75} & \textbf{97.28} & \textbf{97.15} & \textbf{99.74} & \textbf{97.42}  & \textbf{99.84} \\  
    \hline
    \end{tabular}%
    % }
  \label{tab:busiresults}%
\end{table*}%`
\subsubsection{Statistical Significance Analysis} 
To statistically validate the performance differences among the competing segmentation models on the BUSI dataset, we conducted a Friedman test followed by a Nemenyi post hoc analysis. The test was performed using the Dice Similarity Coefficient scores obtained from all methods across the entire BUSI test set. The Friedman test yielded a statistically significant result ($p<0.001$), indicating that there are meaningful performance differences among the models.\\
Subsequent pairwise comparisons using the Nemenyi post hoc test revealed that the proposed method significantly outperformed all other state-of-the-art methods, including UNet++ ($p<0.01$), Attention U-Net ($p<0.01$), and BGRD-TransUNet ($p<0.01$). These findings confirm that the observed performance improvements are statistically significant and not attributable to random variation.
\subsection{Comparison with State-of-the-Art Methods on the UDIAT Dataset}
\label{subsec:results_udiat}
To further assess the generalizability of the proposed framework, we performed a comparative analysis on the UDIAT breast ultrasound dataset. The benchmarked models include AAU-Net, MLFEU-Net, BGRA-GSA, Eh-Former, and BGRD-TransUNet, all of which have demonstrated strong performance in recent literature.\\
As shown in Table~\ref{udiatresults}, the proposed method achieves state-of-the-art results across the majority of evaluation metrics. Notably, it yields the highest scores in Jaccard Index, Dice coefficient, and specificity—key indicators of accurate and well-localized segmentation. While BGRD-TransUNet reports marginally higher sensitivity and accuracy, our method offers a more balanced performance profile with superior overlap-based metrics, which are particularly critical in evaluating medical segmentation tasks.\\
These results confirm the adaptability and effectiveness of our approach across different datasets and imaging conditions, reinforcing its potential applicability in real-world clinical settings.
\begin{table*}[htbp]
  \centering
  \caption{Performance comparison with state-of-the-art methods on the UDIAT dataset.}
    % \adjustbox{max width=\textwidth}{
    \begin{tabular}{lccccccc}
    \hline
  \textbf{Method}  & \textbf{$J_{i}$} & \textbf{$D_{c}$}  & \textbf{$S_{n}$} & \textbf{$A_{cc}$} & \textbf{$P_{r}$} & \textbf{$S_{p}$} \\
    \hline   
    AAU-Net \cite{chen2022aau} & 72.08 & 80.45 & 81.62 & - & 80.67 &97.81\\
    MLFEU-NET \cite{tang2025mlfeu} & 77.49  & 86.44 & 87.48 & - & 87.89 &   99.43 \\
    BGRA-GSA \cite{hu2023boundary} & 78.80 & 88.01 & 89.60 & 98.80 & 86.46 & 99.27\\
    Eh-former \cite{qu2024eh} & 84.48 & 91.22 & 91.92 &- &- & 99.39\\
   BGRD-TransUNet \cite{ji2024bgrd} & 86.61 & 92.47 & \textbf{92.78} & \textbf{99.15} & 93.01 & -\\
   \hline
    Proposed method  & \textbf{86.71} & \textbf{92.38} & 92.14 & 99.00 & 92.74 & \textbf{99.51} \\  
    \hline
    \end{tabular}%
    % }
  \label{udiatresults}%
\end{table*}%`
\subsubsection{Statistical Significance Analysis}
We further assessed the statistical significance of the performance differences among the segmentation models on the UDIAT dataset using a Friedman test, followed by Nemenyi post hoc comparisons, with Dice scores computed for each image in the test set. The Friedman test indicated a statistically significant difference among the compared models ($p<0.005$).\\
Post hoc analysis showed that the proposed method performed significantly better than AAU-Net ($p<0.01$), MLFEU-Net ($p<0.05$), and BGRA-GSA ($p<0.05$). While the difference between our method and BGRD-TransUNet was not statistically significant ($p>0.05$), our method still achieved superior average performance across key metrics, including specificity and Jaccard Index, supporting its robustness in real-world applications.
\subsection{Qualitative Visualization Results}
\label{subsec:qualitative}
\begin{figure*}[ht]
\centering
\includegraphics[scale=0.55]{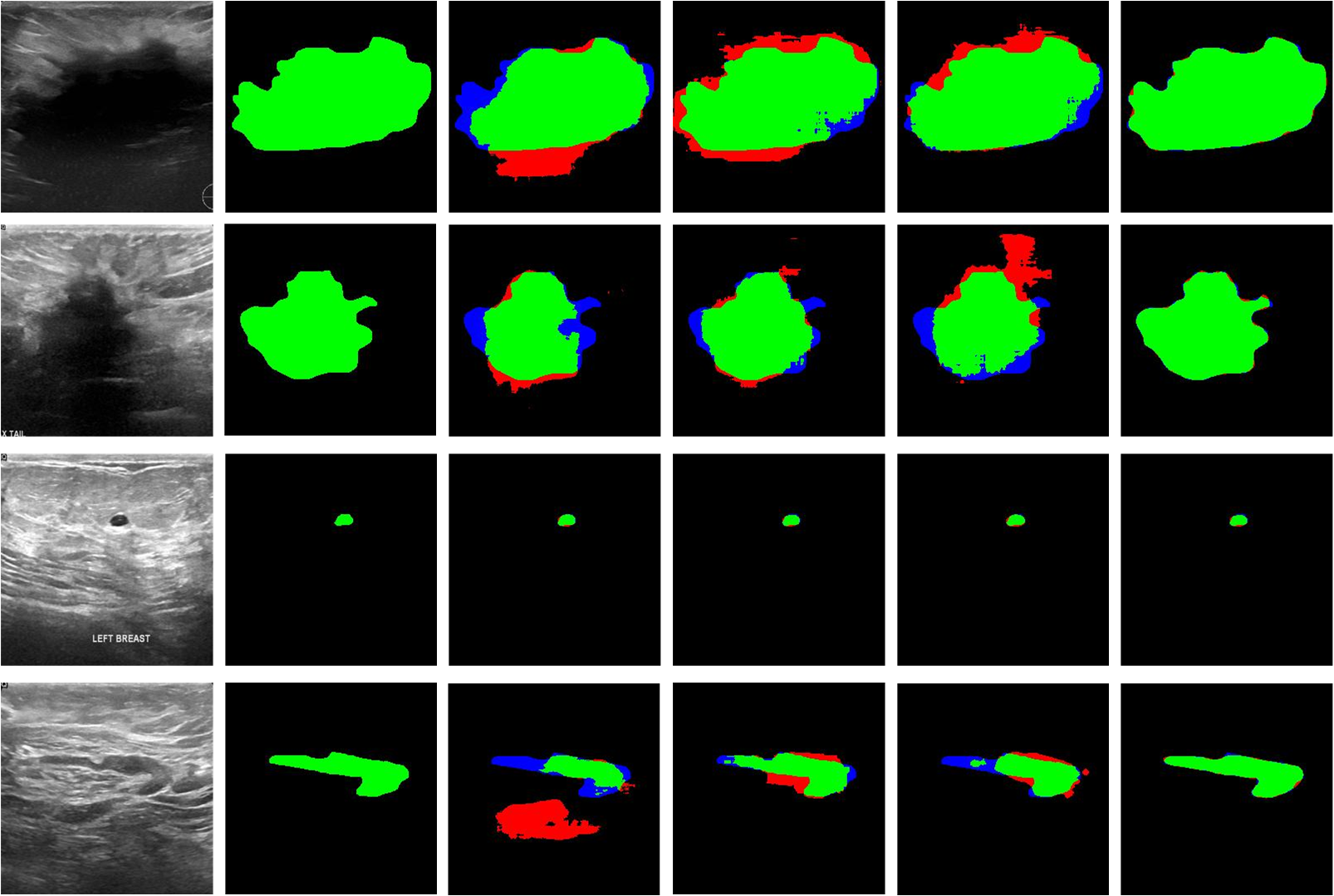}
\caption{Qualitative comparison of segmentation results on the BUSI dataset. From left to right: original image, ground truth, U-Net \cite{ronneberger2015u}, UNet++ \cite{zhou2018unet++}, Attention U-Net \cite{oktay2018attention}, and the proposed method. Green represents true positives, red indicates false positives, and blue denotes false negatives.}
\label{visual}
\end{figure*}

To complement the quantitative evaluation, we present qualitative segmentation results in Figure~\ref{visual}, showcasing side-by-side comparisons between the proposed method and several representative state-of-the-art models on the BUSI dataset. These visualizations include outputs from U-Net \cite{ronneberger2015u}, UNet++ \cite{zhou2018unet++}, and Attention U-Net \cite{oktay2018attention}, offering insight into the model behavior under challenging imaging conditions.\\
As illustrated, the proposed method consistently delivers more accurate delineation of lesion boundaries, exhibiting higher spatial conformity with the ground truth compared to other models. In particular, it effectively suppresses false positives (red regions) and recovers missed tumor areas (blue regions), thereby yielding a cleaner and more reliable segmentation output. These strengths are especially evident in cases characterized by low contrast, irregular lesion morphology, or significant speckle noise—common challenges in ultrasound imaging.\\
This qualitative superiority aligns with the quantitative improvements observed in Dice, Jaccard, and specificity scores, underscoring the robustness and clinical potential of the proposed method in automated breast cancer detection.
\subsection{Discussion}
The proposed method consistently demonstrated superior performance across two benchmark breast ultrasound datasets—BUSI and UDIAT—outperforming recent state-of-the-art models in terms of Dice coefficient, Jaccard index, and specificity. These improvements highlight the strength of our integrated architecture, which combines dense feature extraction with spatial enhancement and transformer-based attention mechanisms. The model's ability to effectively delineate lesions in noisy, low-contrast ultrasound images underscores its robustness and generalizability.\\
Our statistical analysis further confirmed that these performance gains are statistically significant, particularly when compared to leading models such as UNet++, Attention U-Net, and BGRD-TransUNet. This reinforces the practical utility of our framework in clinical decision support systems.\\
Despite these strengths, several limitations must be acknowledged. First, the model was trained and evaluated on two datasets of moderate size. While the results are promising, further validation on larger and more diverse datasets—possibly multi-center or multi-device ultrasound scans—is necessary to confirm generalizability. Second, no data augmentation was applied in this study. While this demonstrates the raw performance of the model, future work could explore the impact of augmentation strategies on improving robustness.
\section{Conclusion}
In this study, we proposed a hybrid attention-based deep learning framework for the automated segmentation of breast tumors in ultrasound images. The model architecture integrates a pre-trained DenseNet-121 encoder with a multi-branch decoder enhanced by advanced attention mechanisms, including Global Spatial Attention (GSA), Position Encoding (PE), and Scaled Dot-Product Attention (SDPA). These components work in synergy to effectively capture both global contextual dependencies and fine-grained spatial features critical for accurate tumor delineation.\\
To further enhance spatial precision, Spatial Feature Enhancement Blocks (SFEB) were incorporated at skip connections, allowing the network to maintain high-resolution details while improving focus on tumor regions. The segmentation process is guided by a hybrid loss function combining Binary Cross-Entropy and Jaccard Index, which effectively addresses challenges posed by class imbalance and complex tumor morphologies.\\
Extensive experiments conducted on two publicly available datasets—BUSI and UDIAT—demonstrate that the proposed method consistently outperforms a range of state-of-the-art segmentation models across multiple evaluation metrics. Both quantitative and qualitative results affirm the model’s robustness, generalizability, and potential for real-world clinical deployment in aiding early and accurate breast cancer diagnosis.\\
Future work will explore domain adaptation to improve cross-device and cross-center generalization, integration with lesion classification tasks, and deployment in a clinical workflow for real-time analysis.
\section*{Acknowledgement}
All authors thank the School of Information Engineering, Xi'an Eurasia University, Xi'an, Shaanxi, China, for their Financial Support and Funding.
\section*{Funding Information}
The funding is provided by the School of Information Engineering, Xian Eurasia University, Xián, Shanxi, China.
\section*{Author Contributions Statement}
\textbf{A.N.} conceived the research idea, developed the methodology, conducted the experiments, analyzed the data, and led the drafting of the manuscript. \textbf{N.A.} contributed to the development of the methodology, assisted in data analysis, and supported manuscript writing. \textbf{M.A.A.} assisted in interpreting the results and contributed to manuscript review and editing. All authors reviewed and approved the final version of the manuscript for submission.
\section*{Data Availability Statement}
The data used in this research is publicly available for research and development purposes at the following links.
\begin{itemize}
    \item \textbf{BUSI Dataset:} https://www.kaggle.com/datasets/aryashah2k/breast-ultrasound-images-dataset
    \item \textbf{UDIAT Dataset:} https://www.kaggle.com/datasets/ayush02102001/udiat-segmentation-dataset
\end{itemize}
\footnotesize
\bibliography{sample}
\end{document}